\documentclass[showpacs,aps,prl,twocolumn,superscriptaddress]{revtex4}
\usepackage{graphicx} 
\usepackage{dcolumn}
\usepackage{bm}
\usepackage{amssymb,amsmath}
\usepackage{epstopdf}

\begin{document}

\title{Renormalized Energies of Superfluorescent Bursts from an Electron-Hole Magneto-plasma with High Gain in InGaAs Quantum Wells}

\normalsize
\author{J.-H.~Kim}
\thanks{J.-H.~Kim and J.~Lee equally contributed to this work.}
\affiliation{Department of Electrical and Computer Engineering, Rice University, Houston, Texas 77005, USA}

\author{J.~Lee}
\thanks{J.-H.~Kim and J.~Lee equally contributed to this work.}
\affiliation{Department of Physics, University of Florida, Gainesville, Florida 32611, USA}

\author{G.~T.~Noe}
\affiliation{Department of Electrical and Computer Engineering, Rice University, Houston, Texas 77005, USA}

\author{Y.~Wang}
\author{A.~K.~W\'{o}jcik}
\affiliation{Department of Physics, Texas A\&M University, College Station, Texas 77843, USA}

\author{S.~A.~McGill}
\affiliation{National High Magnetic Field Laboratory, Tallahassee, Florida 32310, USA}



\author{D.~H.~Reitze}
\affiliation{Department of Physics, University of Florida, Gainesville, Florida 32611, USA}

\author{A.~A.~Belyanin}
\affiliation{Department of Physics, Texas A\&M University, College Station, Texas 77843, USA}

\author{J.~Kono}
\email[]{kono@rice.edu}
\thanks{corresponding author.}
\affiliation{Department of Electrical and Computer Engineering, Rice University, Houston, Texas 77005, USA}
\affiliation{Department of Physics and Astronomy, Rice University, Houston, Texas 77005, USA}

\date{\today}

\begin{abstract}
We study light emission properties of a population-inverted 2D electron-hole plasma in a quantizing magnetic field.  We observe a series of superfluorescent bursts, discrete both in time and energy, corresponding to the cooperative recombination of electron-hole pairs from different Landau levels.  The emission energies are strongly renormalized due to many-body interactions among the photogenerated carriers, exhibiting red-shifts as large as 20\,meV at 15\,T.  However, the magnetic field dependence of the lowest Landau level emission line remains excitonic at all magnetic fields.  Interestingly, our time-resolved measurements show that this lowest-energy burst occurs only after all upper states become empty, suggesting that this excitonic stability is related to the `hidden symmetry' of 2D magneto-excitons expected in the magnetic quantum limit.
\end{abstract}  \maketitle


Optically created electron-hole ($e$-$h$) pairs in semiconductors provide a rich system for the study of carrier interaction effects in a highly controllable manner~\cite{HakenNikitine75Book,HaugKoch93Book,KochetAl06NM}.  With continuously tunable pair density, temperature, and magnetic field, one can systematically examine spectral features of bound and unbound carriers in different regimes.  In the regimes of high densities and/or low temperatures, a variety of many-body ground states can exist~\cite{HalperinRice68RMP,Nozieres83Physica,Keldysh95Chapter,MoskalenkoSnoke00Book}, including Bose-Einstein condensates, $e$-$h$ droplets, and excitonic insulators and crystals.  However, excitons are stable only in the dilute limit, where their Bohr radius is much smaller than the interexciton distance.  As the latter approaches the former, a Mott transition~\cite{Mott61PM} is expected to occur, transforming the insulating excitonic gas into a metallic $e$-$h$ plasma.  In addition, carrier interactions are expected to induce exciton screening and ionization, band-gap renormalization (BGR), and biexciton formation.  Furthermore, optical gain develops, and coherent and cooperative processes such as superradiance (SR) and superfluorescence (SF)~\cite{Dicke1954,BonifacioLuiato1975,BelyaninetAl97QSO,BelyaninetAl98QSO} can dominate the emission spectra.  

\begin{figure}
\includegraphics[scale=0.5] {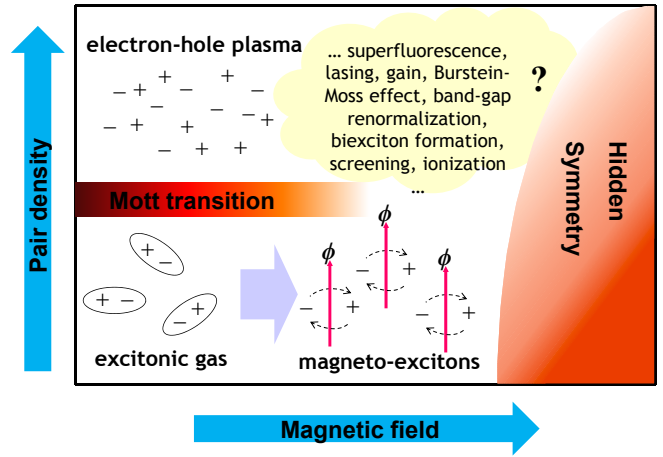}
\caption{(color online). Different phases of electron-hole ($e$-$h$) pairs in a charge neutral semiconductor quantum well as a function of pair density and magnetic field.  At zero field, an insulating excitonic gas transforms into a metallic $e$-$h$ plasma via a Mott transition.  In the high magnetic field limit, `hidden symmetry'~\cite{DzyubenkoLozovik91JPA,ApalkonRashba91JETP,MacDonaldetAl92PRL} is expected to prevent excitons from dissociation into free $e$-$h$ pairs.  
\label{phase}}
\end{figure}

Here, we studied light emission from highly photo-excited undoped InGaAs quantum wells (QWs) at low temperatures and high magnetic fields.  Using intense femtosecond laser pulses, we excited $e$-$h$ pairs at ultrahigh densities and observed bright and sharp emission lines~\cite{JhoetAl06PRL,JhoetAl10PRB}.  At the highest field and excitation ranges, strong Landau quantization and Fermi degeneracy increased the density of states and suppressed decoherence, leading to the emergence of intense and coherent pulses of light, or SF bursts, as we recently reported~\cite{NoeetAl12NP}.  In the present study, we examined the influence of many-body interactions on the light emission processes of $e$-$h$ pairs in the presence of gain, through a detailed magnetic field ($B$) dependent study of the energies of the SF lines.  At zero field, the exciton binding energy approaches zero as the $e$-$h$ pair density increases, i.e., an excitonic Mott transition.  At finite magnetic fields, all lines are strongly red-shifted ($\gtrsim$20\,meV at 15\,T) from the unperturbed energies determined by linear magneto-absorption spectroscopy.  However, the $B$ dependence of the emission line associated with the lowest-energy [or (00)] Landau level remains {\em excitonic}.  At the same time, our time-resolved photoluminescence measurements firmly establish the fact that the (00) SF burst occurs {\em only after all higher-energy SF bursts occur}.  Therefore, we interpret this intriguing stability of the excitonic nature of the (00) SF in terms of `hidden symmetry' of 2D $e$-$h$ systems in a strong perpendicular $B$.  Namely, theoretical studies have proven that there exists a hidden symmetry~\cite{DzyubenkoLozovik91JPA,ApalkonRashba91JETP,MacDonaldetAl92PRL,FinkelsteinetAl97PRB,RashbaetAl00SSC} for $\nu_{e}, \nu_{h} \leq 2$ (where $\nu_{e}, \nu_{h}$ are the electron and hole filling factors), leading to an exact cancelation of all direct and exchange Coulomb interactions among excitons.  This is a consequence of the charge-symmetric nature of $e$-$h$ Coulomb interactions, and as a result, the emission energies are independent of interactions in the high $B$ limit (Fig.~\ref{phase}).

All measurements were performed in the Fast Optics Facility of the National High Magnetic Field Laboratory in Tallahassee, Florida. An undoped multiple quantum well (QW), consisting of fifteen 8-nm In$_{0.2}$Ga$_{0.8}$As QWs and 15-nm GaAs barriers, was placed in a 17.5-T superconducting magnet at 5\,K.  The Mott transition for our sample (effective Bohr radius $a_{\rm B}^*$ $\sim$ 10\,nm) is expected to occur at a density of $(a_{\rm B}^*)^{-2}$ $\sim$ 10$^{12}$~cm$^{-2}$.  Photoluminescence (PL) and transmission spectra were collected through 0.6-mm-core-diameter multimode optical fibers placed at the center and edge of the sample.  A right angle micro-prism (1\,mm$^2$ area) redirected the in-plane emission from the edge of the sample to the fiber~\cite{JhoetAl06PRL,JhoetAl10PRB,NoeetAl12NP}.  A quartz-tungsten-halogen lamp was used for white-light transmission measurements.  For low and high excitation PL studies, a 25-mW He-Ne laser and an amplified Ti:Sapphire laser were used, respectively.  The Ti:Sapphire source produced pulses centered at 800\,nm with a pulse-width of 150\,fs at a 1\,kHz repetition rate.  Pulse energies up to 20\,$\mu$J were used.  When focused to a 0.5\,mm diameter at the sample, pulse fluences of 10\,mJ/cm$^{2}$ were obtained, generating $e$-$h$ pair densities approaching $10^{14}$~cm$^{-2}$.  Time-resolved PL measurements were performed using a streak camera, and details of the setup are fully described in \cite{NoeetAl12NP}.  All measurements were performed in the Faraday geometry, i.e., the $B$ was perpendicular to the QW plane and parallel to the light propagation direction.

\begin{figure}
\includegraphics[scale=0.36] {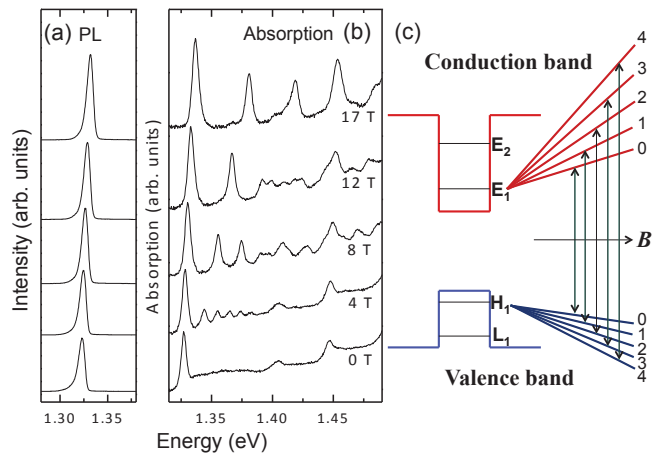}
\caption{Evolution of emission and absorption spectra of InGaAs quantum wells with increasing magnetic field ($B$) in the low density limit.  (a)~Emission spectra under weak He-Ne laser excitation, showing a single peak with a slight diamagnetic shift with $B$.  (b)~Linear absorption spectra taken with a quartz-tungsten-halogen lamp.  The zero-field spectrum consists of a series of step functions superimposed by excitonic peaks at the singularities.  (c)~Schematic energy level diagram with allowed interband magneto-optical transitions between the $E_1$ and $H_1$ subbands indicated by arrows.\label{spectra}}
\end{figure}

Figures \ref{spectra}(a) and \ref{spectra}(b) show PL and absorption spectra, respectively, taken under weak excitation conditions at different $B$. At $B$ = 0, the PL spectrum consists of a single peak from the band-edge exciton, whereas the absorption spectrum consists of a series of step functions typical of a quasi-2D system, superimposed with excitonic peaks at the singularities.  PL shows a single peak at all fields, as shown in Fig.~\ref{spectra}(a), exhibiting a characteristic diamagnetic shift ($\propto B^2$). At finite $B$, the absorption spectrum splits into a series of peaks due to Landau quantization of the conduction and valence bands.  In this study, we consider only states within the first electron and heavy-hole subband ($E_{1}H_{1}$).  We employ the high-field Landau notation ($N,M$), where $N$, $M$ = 0, 1, 2, \ldots, to specify each 2D $e$-$h$ state in a $B$, as opposed to the low-field excitonic notation ($n$,$m$).  The correspondence between the two is given by $n$ = $\max(N,M)$ + 1 and $m$ = $N - M$~\cite{MacDonaldRitchie86PRB}.  Only states with $N$ = $M$ [or, equivalently, $s$-like ($m$ = 0) states] are optically active.

\begin{figure}
\includegraphics[scale=0.55] {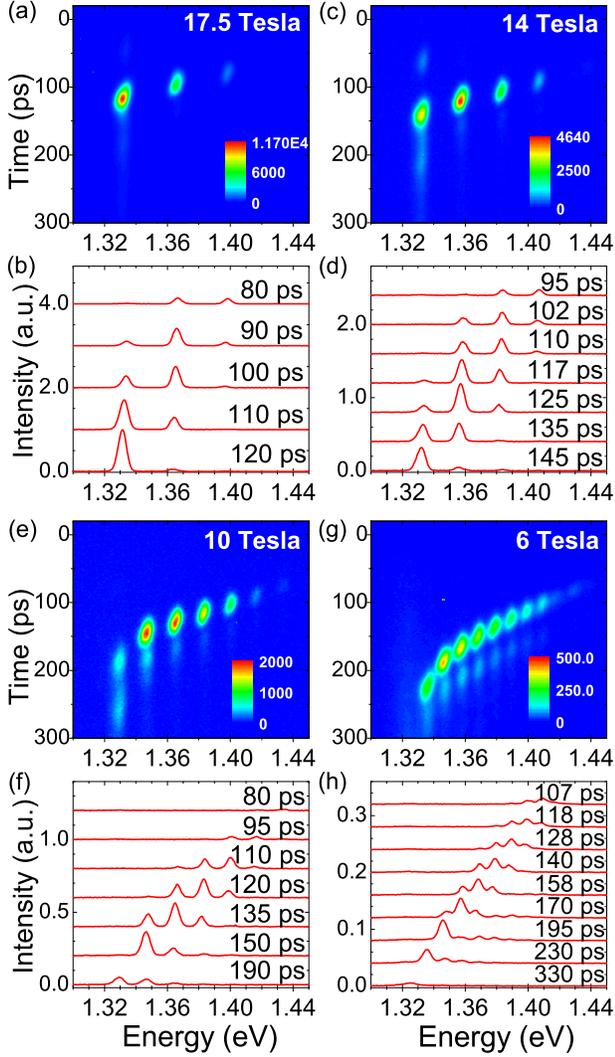}
\caption{(color online). Time-resolved emission spectra at 17.5\,T [(a) and (b)], 14\,T [(c) and (d)], 10\,T [(e) and (f)], and 6\,T [(g) and (h)].  Each ($N$,$N$) recombination is observed as a delayed pulse, or burst, of superfluorescence (SF)~\cite{NoeetAl12NP}.  The delay between the pump pulse and the SF burst increases with decreasing $B$, and at a given $B$, the delay is longer for smaller $N$; the $N$ = 0 state is the last to burst. \label{spectra2}}
\end{figure}

Figures \ref{spectra2}(a)-\ref{spectra2}(h) show time-resolved PL spectra with high excitation using the amplified Ti:Sapphire laser, taken with a streak camera at magnetic fields of 17.5\,T [(a) and (b)], 14\,T [(c) and (d)], 10\,T [(e) and (f)], and 6\,T [(g) and (h)].  Each ($N$,$N$) recombination is observed as a peak both in energy and time, characteristic of a delayed pulse, or burst, of SF~\cite{NoeetAl12NP}.  The delay between the pump pulse and the SF burst changes with $B$, increasing with decreasing $B$.  For the $N$ = 0 state, the delay time is $\sim$120\,ps at 17.5\,T but is $\sim$200\,ps at 10\,T.  Furthermore, at a given $B$, the delay is longer for smaller $N$.  Namely, $e$-$h$ pairs in higher-lying states burst first, and lower-energy states burst in a sequential manner, the $N$ = 0 state being the last to burst.

The $B$-dependence of the energies of the absorption peaks [in Fig.~\ref{spectra}(b)] can be calculated by solving the Schr{\"o}dinger equation for a system including an electron and a heavy hole. The Hamiltonian includes their QW confinement in the growth direction, their interaction  with $B$, and the Coulomb interaction between them. A general eigenstate of the Hamiltonian includes mixing among different levels in the growth direction. However, for the first few eigenstates of the Hamiltonian, this kind of mixing is very small, due to the large energy spacing between them. Thus, we can assume that both the electron and hole are in their ground states described by wave functions $\phi_{e1}(z_e)$ and $\phi_{h1}(z_h)$ in the growth direction, and their in-plane motion experiences an effective Coulomb interaction given by \cite{Ivchenko05Book}
\begin{eqnarray}
V_{\rm eff}(\rho) = - \frac{e^2}{\epsilon} \int d z_e \int d z_h \frac{|\phi_{e1}(z_e)|^2 |\phi_{h1}(z_h)|^2}{\sqrt{\rho^2 + (z_e-z_h)^2}}
\label{Coulomb2Deff}
\end{eqnarray}
where $\rho = |\vec{\rho}_e - \vec{\rho}_h|$ is the in-plane $e$-$h$ separation.

\begin{figure}
\includegraphics[scale=0.55] {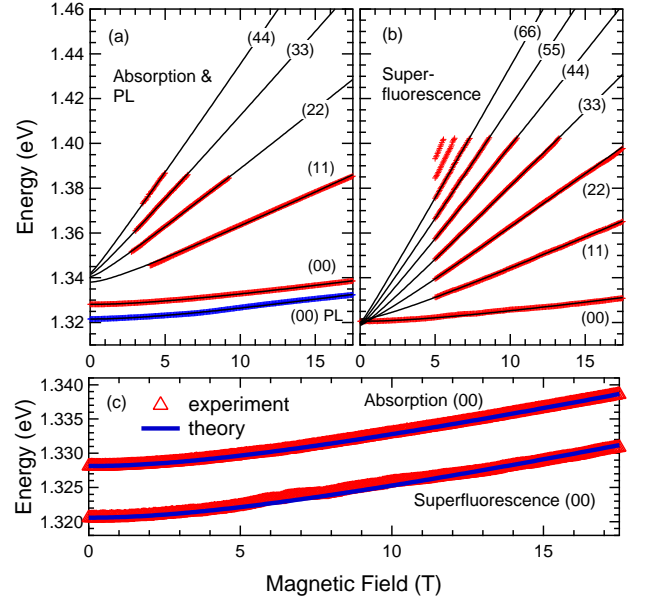}
\caption{(color online). Magnetic field ($B$) dependence of transition energies of observed absorption and emission peaks.  (a)~Low intensity white light absorption (red) and photoluminescence (blue) transition energies versus $B$, together with fits using calculations (solid lines) described in the text.  The Landau indices ($N$,$M$) of each transition are indicated.  The $y$-intercepts at zero field provide the ground and excited excitonic energies with a binding energy of 13.6~meV and band gap of 1.341~eV.  (b)~Transition energies of superfluorescent bursts under high excitation versus $B$, together with calculations (solid lines).  It is seen that all transitions converge to a single point at zero field, indicating a vanishing exciton binding energy at $B = 0$. All transition energies are renormalized with respect to those in (a).  However, the $B$ dependence of the (00) peak completely retains its excitonic character.  (c)~Expanded view of the energy of the (00) line versus $B$ for both absorption and superfluorescence.\label{fit}}
\end{figure}

Using the symmetric gauge $\vec{A}_{e(h)} = \vec{B} \times \vec{\rho}_{e(h)}/2$, we transform the in-plane $e$-$h$ wavefunction $\Psi(\vec{\rho}_e,\vec{\rho}_h)$ as $U(\vec{\rho})$ = $\exp[-i (e / 2 c \hbar) \vec{B} \cdot (\vec{\rho}_{\rm cm} \times \vec{\rho})] \Psi(\vec{\rho}_e,\vec{\rho}_h)$~\cite{ShinadaTanaka70JPSJ}, where we have set the center-of-mass momentum to be zero, since only these states are relevant to optical transitions. If one writes $U(\vec{\rho})$ = $R_m(\rho) e^{i m \phi} / \sqrt{2\pi}$, where $m$ is the angular momentum number, $R_m(\rho)$ satisfies the equation
\begin{eqnarray}
\left\{ -\frac{\hbar^2}{2\mu} \left[ \frac{1}{\rho} \frac{d}{d\rho} \left(\rho \frac{d}{d\rho}\right) - \frac{m^2}{\rho^2} \right] + V_{\rm eff}(\rho) + \frac{\mu \omega_c^2 \rho^2}{8} \right.  \nonumber \\
\left. - \frac{m_e-m_h}{m_e+m_h} \frac{m\hbar\omega_c}{2} \right\} R_m(\rho) = \left(E_m - E_{\rm g,eff}\right) R_m(\rho)
\label{EigenEq}
\end{eqnarray}
where $\mu$ is the reduced mass, $\omega_c = e B/\mu c$, and $E_{\rm g,eff}$ is the effective band gap, which includes the band gap of InGaAs and the ground-level confinement energies of the electron and hole. The boundary conditions are $R_m(0)$ being finite and $R_m(\infty) = 0$. This equation has the Sturm-Liouville form, but is singular at $\rho = 0$ if $m \neq 0$. It can be solved numerically by using the SLEIGN2 code~\cite{BaileyetAl01ACM}. We calculated the eigen-energies for optical transitions ($m$ = 0), and the results agree with the experimental data extremely well, with deviation less than 1\,meV [see Figs.~\ref{fit}(a) and (c)].  Here we assumed fixed values of the dielectric constant $\epsilon = 13.5$ and hole mass $m_h = 0.34m_0$ for all levels but let the electron mass slightly increase with level number to account for the conduction band non-parabolicity: $m_e$ = \{0.0560, 0.0578, 0.0610, 0.0612, 0.0615, 0.0618\} in units of free electron mass $m_0$ for the (00), (11), (22), (33), and (44) transitions, respectively.  The exciton binding energy $E_b$ is $\sim$14\,meV at $B = 0$.

We applied the same procedure to the SF emission data at high densities, except that in this case we treated the effective band gap as a fitting parameter to allow for its renormalization.  As shown in Fig.~\ref{fit}(b), the $B$-dependence of all emission lines is again well fitted, but with renormalized (or enhanced) masses: $m_e$ = \{0.0575, 0.0666, 0.0689, 0.0709, 0.0734, 0.0754, 0.0769\}$m_0$ for the (00), (11), (22), (33), (44), (55), and (66) transitions, respectively.  All transition energies are strongly renormalized with respect to those found in absorption shown in Fig.~\ref{fit}(a); at 15 T, a red-shift of $\gtrsim$20\,meV is observed.  All transition energies converge to a common value at $B$ = 0, indicating the zero-field exciton binding energy $E_b$ = 0 within the uncertainty of fitting ($\pm$~3~meV).  This fact provides clear evidence for an excitonic Mott transition at $B$ = 0, i.e., the high-density $e$-$h$ system at $B$ = 0 is a metallic, $e$-$h$ plasma.  Strikingly, however, the $B$-dependence of the $N$ = 0 peak still exhibits a clear $B^2$ excitonic behavior, as opposed to a linear $B$-dependence indicative of a free $e$-$h$ magnetoplasma. The data agree with our solution for a single magneto-exciton problem within $\pm$~1~meV [see Fig.~\ref{fit}(c)]. 

To interpret these results, we first note the fact that the emission from the Landau levels occurs in a cascaded fashion. In particular, the SF burst from the lowest-energy level, (00), occurs only after all upper states become significantly depopulated by their own bursts of emission, as shown in Figures \ref{spectra2}(a)-\ref{spectra2}(h).  Thus, this amazing stability of excitonic nature at high densities is qualitatively consistent with the concept of the hidden symmetry predicted for 2D magneto-excitons in the high-$B$ limit~\cite{DzyubenkoLozovik91JPA,ApalkonRashba91JETP}, in which all inter-exciton interactions are expected to exactly cancel, thereby immunizing their emission lines against shifting or broadening.  This is a consequence of the charge-symmetric nature of $e$-$h$ Coulomb interactions~\cite{DzyubenkoLozovik91JPA} and has been shown to play an important role in the shake-up process of the lowest Landau level in modulation-doped QWs in high $B$~\cite{FinkelsteinetAl97PRB,RashbaetAl00SSC}.  However, the ranges of $B$ and density in which stability is observed here go beyond the ranges where hidden symmetry has been shown to be valid.  Specifically, hidden symmetry is expected to be valid only in the high-$B$ limit (filling factor $\ll$ 1), where all inter-Landau-level mixing can be ignored, whereas in our studies many higher Landau levels are close in energy to the (00) state and still carry some population by the time of the SF burst. 

The difference of our experiment from previous studies is that we observe not a spontaneous PL but the stimulated emission from a high-gain system. In this case the peak in the stimulated emission spectrum corresponds to the wavenumbers $k$ at which the gain (imaginary part of the frequency Im[$\omega(k)$] of unstable electromagnetic modes) has a maximum.  The remarkable single-exciton behavior of the emission spectrum is a manifestation of the fact that the gain can still be described by the Elliott-type formula~\cite{HaugKoch93Book,KochetAl06NM}, even at high $e$-$h$ densities ($\sim 10^{12}$~cm$^{-2}$) corresponding to a completely filled (00) level. Namely, the gain is dramatically enhanced when the recombining electron and hole spatially overlap forming a neutral two-particle exciton state. Even though the initial spontaneous emission is spread over a broader range of energies corresponding to the transitions between many-particle states of the system, the stimulated emission is concentrated within a narrow spectral region near the peak of the gain at the single-exciton energy. This excitonic character of stimulated emission is extremely robust: it is observed despite the fact that zero-field energies and carrier masses are renormalized at high densities and in spite of rapid changes in the $e$-$h$ populations during SF decay [Fig.~\ref{spectra2}(b)]. Previous studies~\cite{RashbaetAl00SSC} showed no trace of excitonic features in the PL spectra until the filling factor became smaller than 2.  

This work was supported by the NSF (through Grants Nos.~DMR-0325474, DMR-1006663, EEC-0540832, and ECCS-0925446).  A portion of this work was performed at the National High Magnetic Field Laboratory, supported by NSF Cooperative Agreement No.~DMR-0084173 and by the State of Florida.  We thank Glenn S.~Solomon for providing us with the sample used in this study and Prof.~L.~V.~Keldysh for illuminating discussions.



\end{document}